\documentclass[%
 reprint,
 amsmath,amssymb,
 aps,
]{revtex4-2}

\usepackage{graphicx}
\usepackage{dcolumn}
\usepackage{bm}

\usepackage{xcolor}

\begin{document}

\preprint{APS/123-QED}

\title{Cumulant mapping as the basis of multi-dimensional spectrometry}

\author{Leszek J. Frasiński}
  \email{l.j.frasinski@imperial.ac.uk}
\affiliation{%
  Department of Physics, Imperial College London, \\
  London SW7 2AZ, United Kingdom
}%

\date{\today} 

\begin{abstract}
  Cumulant mapping employs a statistical reconstruction of the whole by sampling its parts. The theory developed in this work formalises and extends \textit{ad hoc} methods of `multi-fold' or `multi-dimensional' covariance mapping. Explicit formulae have been derived for the expected values of up to the 6\textsuperscript{th} cumulant and the variance has been calculated for up to the 4\textsuperscript{th} cumulant. A method of extending these formulae to higher cumulants has been described. The formulae take into account reduced fragment detection efficiency and a background from uncorrelated events. Number of samples needed for suppressing the statistical noise to a required level can be estimated using Matlab code included in Supplemental Material. The theory can be used to assess the experimental feasibility of studying molecular fragmentations induced by femtosecond or x-ray free-electron lasers. It is also relevant for extending the conventional mass spectrometry of biomolecules to multiple dimensions. 
\end{abstract}

\maketitle


\section{Introduction and motivation}

Cumulant mapping is an extension of covariance mapping \cite{frasinski1989covariance} to more than two correlated variables. The covariance mapping technique in turn is an extension of the coincidence method \cite{eland1986photoelectron, frasinski1986dissociative} to high counting rates, where several fragmentation events may occur in an elementary sample.

From its invention covariance mapping has been used mostly in studies of ionization and fragmentation of small molecules, with some notable exceptions, such as x-ray scattering or brain studies \cite[see reviews][]{frasinski2016covariance, vallance2021covariance}. Since covariance mapping requires extensive data processing, two-dimensional maps have been most practical. With continued progress in computational power an extension of covariance mapping to higher dimensions is a timely proposition.

Two recent developments have motivated this work. One is a successful application of covariance mapping to two-dimensional mass spectrometry of large biomolecules \cite{driver2020two, driver2021two}, with good prospects for extending this technique to higher dimensions. The second development is the emergence of x-ray free-electron lasers (XFELs), which are powerful research tools for studying atomic and molecular dynamics on the femtosecond and attosecond timescales \cite{huang2021features}. The unprecedented intensity of x-rays in the XFEL pulses induces a large number of fragmentation events, which leaves covariance mapping as the only practical method for correlating the fragments. Moreover, recent XFEL upgrades to high repetition rates, including fast data acquisition, \cite{huang2021features} make the extension of covariance mapping to higher dimensions feasible.

\section{Fragmentation scenario}

\begin{figure}[t]  
  \includegraphics[width=7cm]{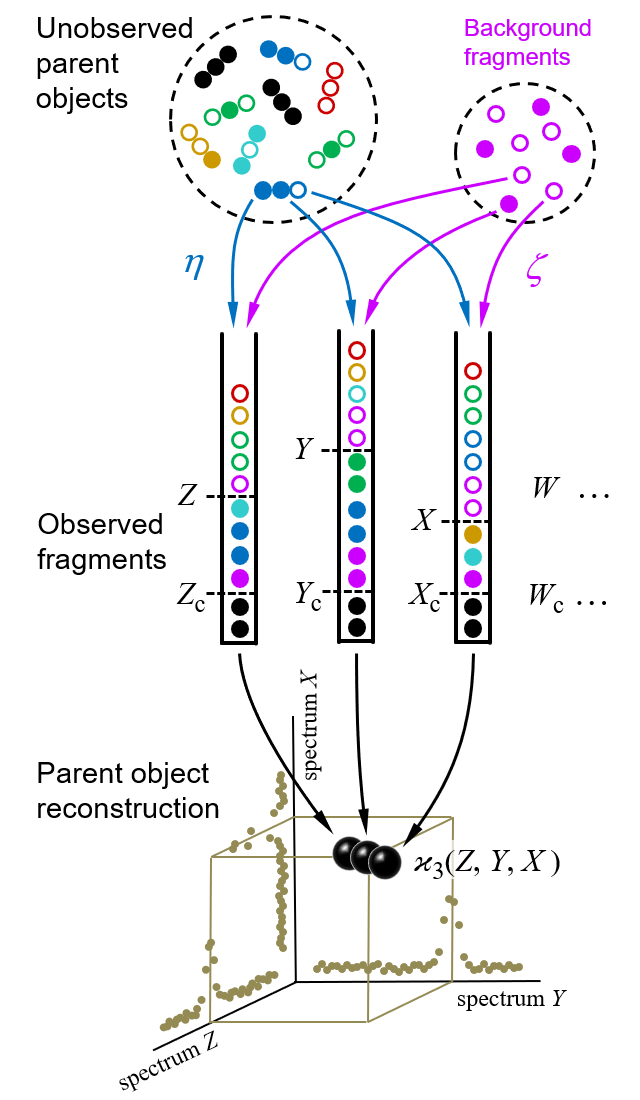}  
  \caption{\label{fig_principle} \textbf{The principle of reconstruction from fragments.} A Poissonian sample of parent objects is fragmented and the fragments are detected with efficiency \(\eta\), including some uncorrelated background in a mean proportion \(\zeta\). The detected fragments (filled circles) are counted and stored in discrete random variables \(Z,Y,X,...\), which are processed over many samples using the cumulant formula. The cumulant value, \(\varkappa_n(Z,Y,X,...)\), measures the number of parent objects at a single point in an \textit{n}-dimensional spectral map.}
\end{figure}

The scheme for cumulant mapping is outlined in Fig.~\ref{fig_principle}. A random sample of unknown objects is drawn from a Poisson distribution. The objects are fragmented, and the fragments are detected. To understand the basic principle, it is helpful to consider initially an ideal scenario where the objects are identical and always break up in the same way into distinguishable fragments. The fragments of each kind are collected in separate bins, and their number is recorded as \(Z, Y, X,\) etc. The sampling is repeated many times and the fragment numbers are used to reconstruct the parent objects.

\subsection{Realistic conditions}

The reconstruction would be trivial under the ideal conditions outlined above. In practice, however, the collected fragment numbers require extensive statistical processing for several reasons.

Firstly, each of the \(Z, Y, X, ...\) random variables is usually measured at just one point on a fluctuating spectrum of mass, energy, or other quantity characterising the fragments, and normally it is not obvious in advance into which parts of the spectra the fragments may fall. Therefore it necessary to calculate the \(\varkappa_n\) cumulant for each possible \textit{n}-tuple of spectral points and display the reconstructed parent objects in an \textit{n}-dimensional map, as shown at the bottom of Fig.~\ref{fig_principle}. Moreover, the spectra of the \((Z, Y, X, ...)\) variables could be more than 1-dimensional, for example, they could be sourced from a position-sensitive timing detector, which effectively resolves fragments in 3D momentum space \cite{allum2021multi}, and in principle would require mapping in 3\textit{n}-dimensional space, unless the fragmentation kinematics can be used to reduce the dimensionality.

Secondly, in most experiments the fragments are detected with quantum efficiency \(\eta < 100\%\). This means that some of the fragments are undetected, as indicated by empty circles in Fig.~\ref{fig_principle}. (The colours mark different detection patterns to enable the reader tracing them from the sample to the bins.) Therefore, the number of tuples \((Z_\text{c}, Y_\text{c}, X_\text{c}, ...)\) of only collectively-correlated detected fragments (black filled circles) is smaller than the number of all detected fragments \(Z, Y, X, ...\) (all filled circles). 

And thirdly, there may be a Poissonian background of fragments (magenta circles) completely unrelated to the sample of interest. This background is characterised by parameter \(\zeta\), which is the ratio of the mean number of the background fragments to the mean number of the sample fragments. Hence, \(\zeta = 0\) means no background, \(\zeta = 1\) means as much background as the sample signal, etc.

Taking into account reduced detection efficiencies and the presence of background fragments relaxes the idealized requirements of identical parent objects and a single fragmentation pattern, which makes cumulant mapping applicable to studies of mixtures and multiplicity of fragmentation channels. The aim of this work is to estimate the cumulant value and noise when the fragments are detected under the non-ideal conditions of \(\eta < 100\%\) and \(\zeta > 0\). 

\subsection{Statistical concepts and notation}

Statistics, like all natural sciences, encompasses two realms: the factual world of tangible measurements and the Platonic world of mathematical abstractions. The objects spanning both worlds are random variables, such as \(Z\), denoted in this work with Roman capitals. 

On one hand, a random variable can be measured yielding sample values, which I denote giving it an index, e.g. \(Z_i\). On the other hand, the general properties of the random variables, such as probability distributions, moments, etc. are mathematical abstractions that cannot be known with absolute certainty. Nevertheless, we can infer these abstract properties by repetitive sampling of the random variable. I assume that in an experiment the conditions stay constant and that we draw \(N\) samples of \(Z\), which are indexed by \(i = 1, 2, 3, ... N\). 

The parameters describing the properties of a random variable are denoted with Greek letters, such as \(\varkappa\). To infer a parameter value we can use the samples to construct an estimator denoted by a hat, e.g. \(\widehat{\varkappa}\). For example, to estimate the first moment of variable \(Z\), we use the sample average  indicated by an overline:
\begin{equation}
  \widehat{\varkappa_1}(Z)  = \overline{Z} = \frac{1}{N} \sum_{i=1}^{N}{Z_i}.
  \label{kappa-1-estim}
\end{equation}

Since repeating the experiment gives us a new set of samples, the sample averages and the estimators are also random variables. However, calculating their expected values (or variance, or higher-order moments) fixes them to the theoretical limit when \(N \rightarrow \infty\). Angular brackets are used to denote the expected values: 
\begin{multline}
  \varkappa_1(Z) = \langle \widehat{\varkappa_1} \rangle
  = \big\langle \overline{Z} \big\rangle
  = \Big\langle \frac{1}{N} \sum_{i=1}^{N}{Z_i} \Big\rangle \\
  = \frac{1}{N} \sum_{i=1}^{N}{\langle Z_i \rangle}
  = \frac{1}{N} N{\langle Z \rangle} = {\langle Z \rangle}.
  \label{kappa-1}
\end{multline}
The bulk of this work is dedicated to such calculations.

\section{Fragment correlations}

In general, \(\widehat{\varkappa_n}\) stands for an estimator of collective correlations among \(n\) fragments. When \(n = 1\) the problem is degenerate and the best we can do is to estimate the mean number of only one kind of a fragment, \(Z\), using the sample average according to Eq.~(\ref{kappa-1-estim}).

\subsection{Covariance}

When \(n = 2\) the appropriate estimator is the sample covariance of the two fragments \(Z\) and \(Y\):
\begin{equation*}
  \widehat{\varkappa_2}(Z,Y)
    = \overline{(Z - \overline{Z})(Y - \overline{Y})}.
\end{equation*}
It is worth noting that that this estimator is biased. In principle, the bias can be removed by using Bessel's correction factor \(N/(N-1)\) each time a degree of freedom of the sample has been used to calculate an inner average. In practice however, the bias is insignificant for \(N \gg 1\) and can be ignored where appropriate.

Calculating the expected value of this estimator leads to the well known formula for covariance:
\begin{multline*}
  \varkappa_2(Z,Y) = \langle \widehat{\varkappa_2} \rangle 
  = \Big\langle \overline{(Z - \overline{Z})(Y - \overline{Y})} \Big\rangle \\
  = \big\langle (Z - \overline{Z})(Y - \overline{Y}) \big\rangle
  = \langle (Z - \langle Z \rangle)(Y - \langle Y \rangle) \rangle \\
  = \langle ZY \rangle - \langle Z \rangle \langle Y \rangle
  = \text{cov}(Z, Y).
\end{multline*}
Introducing mean-centered variables
\begin{equation}
  z_0 = Z -\langle Z \rangle, \;\; y_0 = Y -\langle Y \rangle ,
  \label{z0-y0}
\end{equation}
gives us a compact version of the formula:
\begin{equation}
  \varkappa_2(Z,Y) = \langle z_0 y_0 \rangle .
  \label{kappa-2}
\end{equation}
(Symbols \(z\), \(y\), \(x\), etc. are reserved for later use.)

\subsection{The problem of more than two fragments}

When there are three fragments, an extension of Eq.~(\ref{kappa-2}) has been proposed \cite{frasinski1991multiphoton}:
\begin{equation}
  \varkappa_3(Z,Y,X) = \langle z_0 y_0 x_0 \rangle ,
  \label{kappa-3}
\end{equation}
and the suitability of this ``3-fold covariance'' formula has been demonstrated experimentally \cite{frasinski1991multiphoton, bryan2006observation} and theoretically \cite{mikosch2013coincidence}.

One may expect that this method of extending the covariance formula works for four fragments:
\[ \varkappa_4^\text{trial}(Z,Y,X,W) = \langle z_0 y_0 x_0 w_0 \rangle. \]
Unfortunately, this trial for the formula is unsuitable \cite{zhaunerchyk2014theory}. The reason is that if we have only pairwise correlations, e.g. \(Z\) with \(Y\) and \(X\) with \(W\), then
\[ \varkappa_4^\text{trial} = \langle z_0 y_0\rangle
  \langle x_0 w_0 \rangle \neq 0,\]
but we want \(\varkappa_4 = 0\) because there is no collective correlation among all four fragments.

\subsection{The solution}

To find the correct formula for \(n \geq 4\), I start with listing the desired properties of \(\varkappa_n = \varkappa_n(Z,Y,X, \ldots)\):
\begin{itemize}
  \item \(\varkappa_n \neq 0\) only if all arguments are collectively correlated;
  \item \(\varkappa_n\) has units of the product of all arguments;
  \item \(\varkappa_n\) is linear in the arguments;
  \item \(\varkappa_n\) is invariant under interchange of any two arguments.
\end{itemize}

It turns out that these properties uniquely determine the formula. For example, the reader is invited to check that the following formula has the desired properties:
\begin{multline*}
  \varkappa_4(Z,Y,X,W) = \langle z_0 y_0 x_0 w_0 \rangle \\
  - (\langle z_0 y_0 \rangle \langle x_0 w_0 \rangle
  + \langle z_0 x_0 \rangle \langle y_0 w_0 \rangle
  + \langle z_0 w_0 \rangle \langle y_0 x_0 \rangle),
\end{multline*}
and that other products of expected values, such as \(\langle z_0 \rangle \langle y_0 x_0 w_0 \rangle\) cannot contribute to the formula because 
\[\langle z_0 \rangle = \langle Z - \langle Z \rangle \rangle
  = \langle Z \rangle - \langle Z \rangle = 0.\]
The formula for \(\varkappa_4\) can be simplified by writing
\begin{equation}
\varkappa_4(Z,Y,X,W) = \langle z_0 y_0 x_0 w_0 \rangle 
  - \sum^3 \langle z_0 y_0 \rangle \langle x_0 w_0 \rangle,
  \label{kappa-4}
\end{equation}
where \(\sum^3\) denotes a sum over all ways of paring the four variables.

Similarly,
\begin{multline}
  \varkappa_5(Z,Y,X,W,V) = \langle z_0 y_0 x_0 w_0 v_0 \rangle \\
  - \sum^{10} \langle z_0 y_0 \rangle \langle x_0 w_0 v_0 \rangle,
  \label{kappa-5}
\end{multline}
and
\begin{multline}
  \varkappa_6(Z,Y,X,W,V,U) = \langle z_0 y_0 x_0 w_0 v_0 u_0 \rangle \\
  - \sum^{15} \langle z_0 y_0 \rangle \langle x_0 w_0 v_0 u_0 \rangle
  - \sum^{10} \langle z_0 y_0 x_0 \rangle \langle w_0 v_0 u_0 \rangle \\
  + 2 \sum^{15} \langle z_0 y_0 \rangle \langle x_0 w_0 \rangle
    \langle v_0 u_0 \rangle.
  \label{kappa-6}
\end{multline}
Formulae for collective correlations among more variables can be constructed in a similar manner.

\subsection{Cumulants}

In statistics the \(\varkappa_n\) parameter that measure the collective correlations of \textit{n} random variables is known as the \textit{n}-variate joint cumulant of the first order \cite[Ref.][chapters 3, 12 and 13]{kendall94vol1}. In this work I shall shorten this name and call it simply the \textit{n}\textsuperscript{th} cumulant.

Cumulants are useful in seemingly disjoint areas of science, such as light--matter interactions \cite{sanchez2020cumulant}, quantum theory of multi-electron correlations \cite{kutzelnigg1999cumulant}, bond breaking of diatomic molecules \cite{brea2013behavior}, neural network theory \cite{helias2020statistical}, financial data analysis \cite{domino2020multivariate}, and gravitational interaction of dark matter \cite{uhlemann2018finding}. In physics multivariate cumulants are also known as the Ursell functions \cite{ursell1927evaluation}.

\subsection{Physical interpretation of  \(\varkappa_n\)}

So far, the cumulants have been introduced as parameters describing multivariate distributions. Now we want to apply them to the problem of recovering the parent objects from their fragments.

Let us suppose that we repetitively gather a sample of objects in a random manner, so the number of objects \(S\) in the sample follows the Poisson distribution
\begin{equation}
  S \sim \text{Pois}(\lambda) \equiv P(S=k)
  = \frac{\lambda^k}{k!}e^{-\lambda},
  \label{pois}
\end{equation}
where \(P(S=k)\) is the probability of having exactly \(k\) objects in the sample and parameter \(\lambda\) is the expected number of the objects in a sample. Next, we fragment the objects and detect only the fragments. And from the detected fragments  we want to infer the identity of the undetected parent objects.

To understand why cumulants are useful in this task, we first consider ideal conditions: there is only one way of fragmenting the parent, we detect every fragment, and there is no background of fragments from other processes. Such fragmentation process can be written as
\[S \rightarrow (Z, Y, X, ...),\]
where \((Z, Y, X, ...)\) is a tuple containing \(n\) fragments. In such a simple process the number of fragments of each kind matches the number of parent objects. Hence, the random variables are equal:
\[Z = Y = X = ... = S\]
and
\[z_0 = y_0 = y_0 = ... = s_0,\]
which reduces the expected values in Eqs. \ref{kappa-1} and \ref{kappa-2}--\ref{kappa-6} to the central moments of the Poisson distribution:
\[\langle z_0 y_0 x_0 ... \rangle  = \langle s_0^n \rangle
  = \langle (S - \langle S \rangle)^n \rangle = \mu_n .\]
Since these moments are known polynomials of \(\lambda\) \cite[Ref.][Section 5.9]{kendall94vol1}, the calculation of cumulants is straightforward:
\begin{subequations}
\begin{align}
  \varkappa_1(S) &= \langle S \rangle = \lambda, \\
  \varkappa_2(S,S) &= \mu_2 = \lambda, \\
  \varkappa_3(S,S,S) &= \mu_3 = \lambda, \\
  \varkappa_4(S,S,S,S) &= \mu_4 - 3\mu_2^2 \nonumber \\
    &= (\lambda + 3\lambda^2) - 3\lambda^2 = \lambda.
\end{align}
  \label{moments}
\end{subequations}
Note how the sum in Eq.~\ref{kappa-4} cancels out the higher powers of \(\lambda\) leaving just the linear term. In fact this is the general property of the Poisson-distribution cumulants \cite[Ref.][Example 3.10]{kendall94vol1}:
\[\varkappa_n(S, S, S, ...) = \lambda.\]

We conclude that under the ideal fragmentation scenario cumulant mapping of fragments gives us a statistical estimate of the mean number of parent objects.

\section{Uncorrelated background}

Ideal conditions are rarely met in practice. We should estimate how a reduced detection efficiency and a background from uncorrelated fragments affect cumulant mapping. Both effects influence the statistics of the cumulant estimator in a similar manner; those fragments that are detected but have undetected siblings effectively contribute to the uncorrelated background, which is depicted in Fig.~\ref{fig_principle} using non-black filled circles.

When \(\eta < 100\%\) or \(\zeta > 0\), then the random variables \(S, Z, Y, ...\) are only partially correlated. To find the formula for \(\varkappa_n\) we need to separate the correlated and uncorrelated parts of these variables.

\subsection{Correlated and uncorrelated parts}

While the parent objects in the sample still follow the Poisson distribution given by Eq.~\ref{pois}, the reduced detection efficiency effectively combines the parent distribution with a binomial distribution of the partial detection giving another Poisson distribution:
\begin{equation*}
  Z \sim \text{Pois}(\lambda) \ast \text{Binom}(\eta_Z)
    \rightarrow \text{Pois}(\eta_Z \lambda).
\end{equation*}
When a Poissonian background from other, uncorrelated fragments is added, the compound probability distribution continues to be Poissonian with a modified expected value:
\begin{equation*}
  Z \sim \text{Pois}(\eta_Z \lambda) \diamond \text{Pois}(\eta_Z \zeta_Z \lambda)
    \rightarrow \text{Pois}(\eta_Z (1 + \zeta_Z) \lambda).
\end{equation*}
Similarly, \(Y \sim \text{Pois}(\eta_Y (1 + \zeta_Y) \lambda)\), etc.

Since the binomial sampling of each of the \(Z, Y, X, ...\) fragments is independent, their joint detection efficiency is a product of the individual efficiencies. Therefore the probability distribution of the detected fragments \(Z\) correlated with all the other detected fragments is given by
\begin{equation*}
  Z_\text{c} \sim \text{Pois}(\theta_n \lambda),
\end{equation*}
where
\[\theta_n = \eta_Z \eta_Y \eta_X ... \; .\]
By the same reasoning \(Y_\text{c} \sim \text{Pois}(\theta_n \lambda), X_\text{c} \sim \text{Pois}(\theta_n \lambda),\) etc. Moreover, the correlated parts are present in each kind of fragment to the same extent, therefore:
\begin{equation}
  Z_\text{c} = Y_\text{c} = X_\text{c} = ...
    = S_\text{c} \sim \text{Pois}(\theta_n \lambda)
  \label{pois-corr}.
\end{equation}

Since the number of detected fragments is the sum of the correlated and uncorrelated parts:
\begin{equation}
  Z = Z_\text{c} + Z_\text{u}, \;\; Y = Y_\text{c} + Y_\text{u}, \;\;
  X = X_\text{c} + X_\text{u}, \;\; ... 
  \label{Z-Y-X}
\end{equation}
and a sum of Poisson distributions is a Poisson distribution, the distributions of uncorrelated parts can be found as follows:
\begin{align*}
  Z_\text{u} &= Z - Z_\text{c} \sim \text{Pois}(\eta_Z (1 + \zeta_Z) \lambda - \theta_n \lambda) \\
             &= \text{Pois}((\eta_Z (1 + \zeta_Z) - \theta_n) \lambda), \\
  Y_\text{u} & \sim \text{Pois}((\eta_Y (1 + \zeta_Y) - \theta_n) \lambda), \\
  X_\text{u} & \sim \text{Pois}((\eta_X (1 + \zeta_X) - \theta_n) \lambda), ...
\end{align*}

Further calculations are significantly simplified when mean-centered variables are introduced for the correlated and uncorrelated parts:
\begin{align}
  s &= S_\text{c} - \langle S_\text{c} \rangle, \nonumber \\
  z &= Z_\text{u} - \langle Z_\text{u} \rangle, \nonumber \\
  y &= Y_\text{u} - \langle Y_\text{u} \rangle, \nonumber \\
  x &= X_\text{u} - \langle X_\text{u} \rangle, \; ...  \; .
  \label{s-z-y-x}
\end{align}
Using Eqs. \ref{z0-y0}, \ref{Z-Y-X}, and \ref{pois-corr} we obtain
\begin{align}
  z_0 &= s + z, \nonumber \\
  y_0 &= s + y, \nonumber \\
  x_0 &= s + x, ... \; .
  \label{z0-y0-x0}
\end{align}

One useful implication of Eq.~\ref{s-z-y-x} is that the expected values of the mean-centered variables vanish:
\begin{equation}
  \langle s \rangle = \langle z \rangle = \langle y \rangle
  = \langle x \rangle = ... = 0.
  \label{vanish}
\end{equation}

The second useful property is that they can be regarded as independent, which is exactly true if only one or two detectable fragments are produced. For three or more fragments it may happen that one fragment that is not detected relegates the other fragments to the background in spite of their correlation. It can be shown that these residual correlations do not affect the expected values of cumulants, nor the variance of the first and second cumulant.

\subsection{Expected values of cumulants}

\begin{figure}[t]  
  \includegraphics[width=7cm]{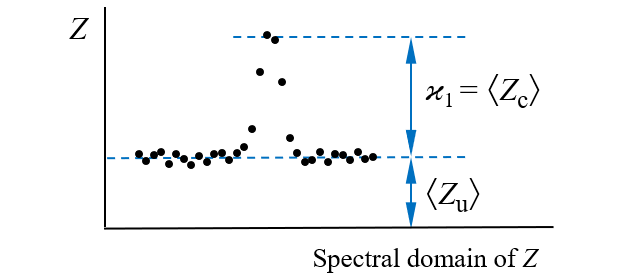}  
  \caption{\label{fig_kappa1} \textbf{How to separate the first cumulant from the uncorrelated background.} Cumulant \(\varkappa_1\) measures only the correlated fragments \(Z_\text{c}\) that form a peak on a spectrum. The peak rides on a background of uncorrelated fragments \(Z_\text{u}\). This definition of \(\varkappa_1\) keeps the background parameter \(\zeta_Z =\langle Z_\text{u} \rangle / \langle Z_\text{c} \rangle\) consistent with the higher cumulants.}
\end{figure}

The first cumulant is unusual because formally it cannot distinguish between the sample fragments and the uncorrelated background. To deal with this ambiguity I shall redefine the first cumulant given by Eq.~\ref{kappa-1}, so it measures only the fragments coming from the sample:
\begin{equation}
  \varkappa_1(Z) = \langle Z_\text{c} \rangle = \langle S_\text{c} \rangle
  = \eta_Z \lambda = \theta_1 \lambda .
  \label{kappa-1-bgnd}
\end{equation}
This definition not only makes \(\varkappa_1\) consistent with the higher cumulants but also gives it the meaning of a signal that is separate from the background. As illustrated in Fig.~\ref{fig_kappa1}, when in spectral analysis the sample fragments form a peak riding on a broad background given by \(\langle Z_\text{u} \rangle = \eta_Z \zeta_Z\lambda\), then \(\langle Z_\text{c} \rangle = \eta_Z \lambda\) is just the peak height.

The higher cumulants can be calculated using the mean-centered variables, \(z_0 = s + z, \; y_0 = s + y, \:\) etc. For example,
\begin{multline}
  \varkappa_2(Z,Y) 
  \overset{\ref{kappa-2}}{=} \langle z_0 y_0 \rangle \
  \overset{\ref{z0-y0-x0}}{=} \langle (s + z)(s + y) \rangle \\
  = \langle s^2 \rangle
  + \langle s \rangle (\langle z \rangle + \langle y \rangle)
  + \langle z \rangle \langle y \rangle \\
  \overset{\ref{vanish}}{=} \langle s^2 \rangle 
  \overset{\ref{s-z-y-x},\ref{pois-corr}}{=} \theta_2 \lambda,
  \label{kappa-2-bgnd}
\end{multline}
where the numbers above equality signs refer to the equations used.

We notice that the formulae for all higher cumulants can be derived in a similar manner. When expanding \(\langle z_0 y_0 x_0 ... \rangle\), most of the terms vanish because of Eq.~\ref{vanish}, and we are left only with polynomials of moments of \(s\). These polynomials are the same as in Eqs.~\ref{moments} except now \(\varkappa_n\) follows \(\text{Pois}(\theta_n \lambda)\) rather then \(\text{Pois}(\lambda)\). Therefore, we obtain a general result:
\begin{equation}
  \varkappa_n(Z,Y,X,...) = \theta_n \lambda.
  \label{kappa-n}
\end{equation}
The simplicity of this result is remarkable. Despite extensive data processing required to estimate cumulants, their meaning is simple: cumulants reconstruct objects from partially detected fragments disregarding any background of other fragments.

\subsection{Estimators of cumulants}

To construct cumulant estimators, the expected values in Eqs.~\ref{kappa-2}--\ref{kappa-6} should be replaced with sample averages. The simplest action is to use Eq.~\ref{kappa-1-estim} everywhere. If, however, unbiased estimators are desired, factor \(1/N\) should be replaced with \(1/(N-1)\) whenever a degree of freedom has already been used, for example
\begin{multline}
  \widehat{\varkappa_2}(Z,Y)
  = \overline{(Z - \overline{Z})(Y - \overline{Y})} \\
  = \frac{1}{N-1} \sum_{i=1}^{N}{
    \Big (Z_i - \frac{1}{N} \sum_{i=j}^{N}{Z_j} \Big )
    \Big (Y_i - \frac{1}{N} \sum_{i=j}^{N}{Y_j} \Big )}.
  \label{estim-2}
\end{multline}
These estimators can be plotted as 2-dimensional maps \cite{frasinski1989covariance} or slices of higher-dimensional maps \cite{frasinski1991multiphoton}.

\subsection{Noise of estimators}

Due to the finite number of samples collected, cumulant estimators are noisy. When assessing the feasibility of an experiment involving a cumulant map, the expected noise on the map is of primary concern. It is known that with increasing dimensionality of the map, the noise-to-signal ratio (N/S) increases \cite{frasinski1991multiphoton}. There are two sources of this deterioration. Firstly, each time the map dimensionality is increased, the signal decreases because it is multiplied by the detection efficiency according to Eq.~\ref{kappa-n}. And secondly, the higher the cumulant, the more subtraction of lower correlations is needed, which contributes more noise from the subtrahends.

These effects can be quantified by calculating the variance of the cumulant estimator, \(\text{var}(\widehat{\varkappa_n})\), and finding the noise-to-signal ratio: 
\begin{equation}
  \text{N/S} = \sigma_n / \varkappa_n \text{, where }
  \sigma_n = \sqrt{\text{var}(\widehat{\varkappa_n})}.
  \label{NS}
\end{equation}
Since the calculations of the variance are lengthy, they are relegated to the appendix. As usual, we find that the standard deviation \(\sigma_n \propto 1/\sqrt{N}\), therefore, once we know \(\sigma_n\) we can estimate the number of samples, \(N\), needed for the required noise-to-signal ratio and assess the experimental feasibility.

\section{Summary of analytical results} \label{summary}

The values and variances of the cumulants are quite complicated functions of the counting rate, \(\lambda\), the detection efficiency, \(\eta\), and the relative background, \(\zeta\). Rather than inspecting the analytical formulae, it is more informative to plot the results for some chosen argument ranges. Matlab code that calculates the values and variances of cumulants up to the 4\textsuperscript{th} one is  included in Supplemental Material. When reading and using the code, it is helpful to refer to the equations written in normal mathematical notation.

The equations give cumulant estimates \(\widehat{\varkappa_n}\) constructed from the samples according to Eq.~\ref{kappa-1-estim}, the cumulant values \(\varkappa_n\), and the cumulant variances \(\text{var}(\widehat{\varkappa_n})\). The auxiliary quantities are the central moments of the correlated parts, \(\langle s^n \rangle\), and of the uncorrelated parts, \(\langle z^2 \rangle\), \(\langle y^2 \rangle\), \(\langle x^2 \rangle\), and \(\langle w^2 \rangle\) (see Eq.~\ref{s-z-y-x}). 

\subsection{1D spectrum} \label{summ1D}

Note that the expected value of the first cumulant depends only on the correlated part, which is justified in the discussion of Eq.~\ref{kappa-1-bgnd}. Hence
\begin{align*}
  \widehat{\varkappa_1} &= \overline{Z} = \overline{Z_\text{c}} + \overline{Z_\text{u}}, \\
  \varkappa_1 &= \big\langle \overline{Z_\text{c}} \big\rangle
     = \langle S_\text{c} \rangle = \theta_1 \lambda, \\
  \text{var}(\widehat{\varkappa_1})
    &= \frac{1}{N} \big( \langle s^2 \rangle + \langle z^2 \rangle \big), \\
\end{align*}
where
\begin{align*}
  \langle s^2 \rangle &= \theta_1 \lambda, \\
  \langle z^2 \rangle &= \big((1 + \zeta_Z) \eta_Z - \theta_1 \big) \lambda, \\
  \theta_1 &= \eta_Z.
\end{align*}

\subsection{2D covariance map} \label{summ2D}

The second cumulant is commonly known as covariance.

\begin{align*}
  \widehat{\varkappa_2} &= \overline{(Z - \overline{Z})(Y - \overline{Y})}, \\
  \varkappa_2 &= \langle \widehat{\varkappa_2} \rangle
     = \langle s^2 \rangle = \theta_2 \lambda, \\
  \text{var}(\widehat{\varkappa_2})
    &\approx \frac{1}{N} \Big( \langle s^4 \rangle - \langle s^2 \rangle^2 \\
    &+ \langle s^2 \rangle 
       \big(\langle z^2 \rangle + \langle y^2 \rangle \big) \\
    &+ \langle z^2 \rangle \langle y^2 \rangle \Big), \\
\end{align*}
where
\begin{align*}
  \langle s^2 \rangle &= \theta_2 \lambda, \\
  \langle s^4 \rangle &= \theta_2 \lambda + 3 \,\theta_2^2 \lambda^2, \\
  \langle z^2 \rangle &= \big((1 + \zeta_Z) \eta_Z - \theta_2 \big) \lambda, \\
  \langle y^2 \rangle &= \big((1 + \zeta_Y) \eta_Y - \theta_2 \big) \lambda, \\
  \theta_2 &= \eta_Z \eta_Y.
\end{align*}

\subsection{3D cumulant map} \label{summ3D}

The third cumulant is sometimes called 3-fold covariance or 3-dimensional covariance.

\begin{align*}
  \widehat{\varkappa_3} &=
    \overline{(Z - \overline{Z})(Y - \overline{Y})(X - \overline{X})}, \\
  \varkappa_3 &= \langle \widehat{\varkappa_3} \rangle
     = \langle s^3 \rangle = \theta_3 \lambda, \\
  \text{var}(\widehat{\varkappa_3})
    &\approx \frac{1}{N} \Big( \langle s^6 \rangle - \langle s^3 \rangle^2 \\
    &+ \langle s^4 \rangle \sum^3 \langle z^2 \rangle \\
    &+ \langle s^2 \rangle \sum^3 \langle z^2 \rangle \langle y^2 \rangle \\
    &+ \langle z^2 \rangle \langle y^2 \rangle \langle x^2 \rangle \Big), \\
\end{align*}
where
\begin{align*}
  \langle s^2 \rangle &= \theta_3 \lambda, \\
  \langle s^3 \rangle &= \theta_3 \lambda, \\
  \langle s^4 \rangle &= \theta_3 \lambda + 3\,\theta_3^2 \lambda^2, \\
  \langle s^6 \rangle &= \theta_3 \lambda + 25\,\theta_3^2 \lambda^2
                       + 15\,\theta_3^3 \lambda^3, \\
  \langle z^2 \rangle &= \big((1 + \zeta_Z) \eta_Z - \theta_3 \big) \lambda, \\
  \langle y^2 \rangle &= \big((1 + \zeta_Y) \eta_Y - \theta_3 \big) \lambda, \\
  \langle x^2 \rangle &= \big((1 + \zeta_X) \eta_X - \theta_3 \big) \lambda, \\
  \theta_3 &= \eta_Z \eta_Y \eta_X.
\end{align*}

\subsection{4D cumulant map} \label{summ4D}

The fourth cumulant formulae are the main result of this work.

\begin{align*}
  \widehat{\varkappa_4} &= \overline{
    (Z - \overline{Z})(Y - \overline{Y})(X - \overline{X})(W - \overline{W})} \\
    &- \sum^3 \overline{(Z - \overline{Z})(Y - \overline{Y})} \:\:
              \overline{(X - \overline{X})(W - \overline{W})}, \\
  \varkappa_4 &= \langle \widehat{\varkappa_4} \rangle 
     = \langle s^4 \rangle - 3 \langle s^2 \rangle^2 = \theta_4 \lambda, \\
  \text{var}(\widehat{\varkappa_4})
    &\approx \frac{1}{N} \Big( \langle s^8 \rangle - \langle s^4 \rangle^2 \\
    &+ 48\langle s^4 \rangle \langle s^2 \rangle^2
     - 12\langle s^6 \rangle \langle s^2 \rangle - 36\langle s^2 \rangle^4 \\
    &+ \big(\langle s^6 \rangle - 6\langle s^4 \rangle \langle s^2 \rangle
     + 9\langle s^2 \rangle^3 \big) \sum^4 \langle z^2 \rangle \\
    &+ \big(\langle s^4 \rangle - \langle s^2 \rangle^2 \big)
       \sum^6 \langle z^2 \rangle \langle y^2 \rangle \\
    &+ \langle s^2 \rangle
        \sum^4 \langle z^2 \rangle \langle y^2 \rangle \langle x^2 \rangle \\
    &+ \langle z^2 \rangle \langle y^2 \rangle
       \langle x^2 \rangle \langle x^2 \rangle \Big), \\
\end{align*}
where
\begin{align*}
  \langle s^2 \rangle &= \theta_4 \lambda, \\
  \langle s^4 \rangle &= \theta_4 \lambda + 3\,\theta_4^2 \lambda^2, \\
  \langle s^6 \rangle &= \theta_4 \lambda + 25\,\theta_4^2 \lambda^2
                       + 15\,\theta_4^3 \lambda^3, \\
  \langle s^8 \rangle &= \theta_4 \lambda + 119\,\theta_4^2 \lambda^2
                    + 490\,\theta_4^3 \lambda^3 + 105\,\theta_4^4 \lambda^4, \\
  \langle z^2 \rangle &= \big((1 + \zeta_Z) \eta_Z - \theta_4 \big) \lambda, \\
  \langle y^2 \rangle &= \big((1 + \zeta_Y) \eta_Y - \theta_4 \big) \lambda, \\
  \langle x^2 \rangle &= \big((1 + \zeta_X) \eta_X - \theta_4 \big) \lambda, \\
  \langle w^2 \rangle &= \big((1 + \zeta_W) \eta_W - \theta_4 \big) \lambda, \\
  \theta_4 &= \eta_Z \eta_Y \eta_X \eta_W.
\end{align*}

\section{Discussion of the results}

The figures in this section have been drawn using the Matlab code given in Supplemental Material. For a detailed inspection of the results, it is recommended to run the code and vary the figure options, such as rotate the 3D plots or change the argument ranges.

\subsection{Ideal conditions}

\begin{figure}[b]  
  \includegraphics[width=\linewidth]{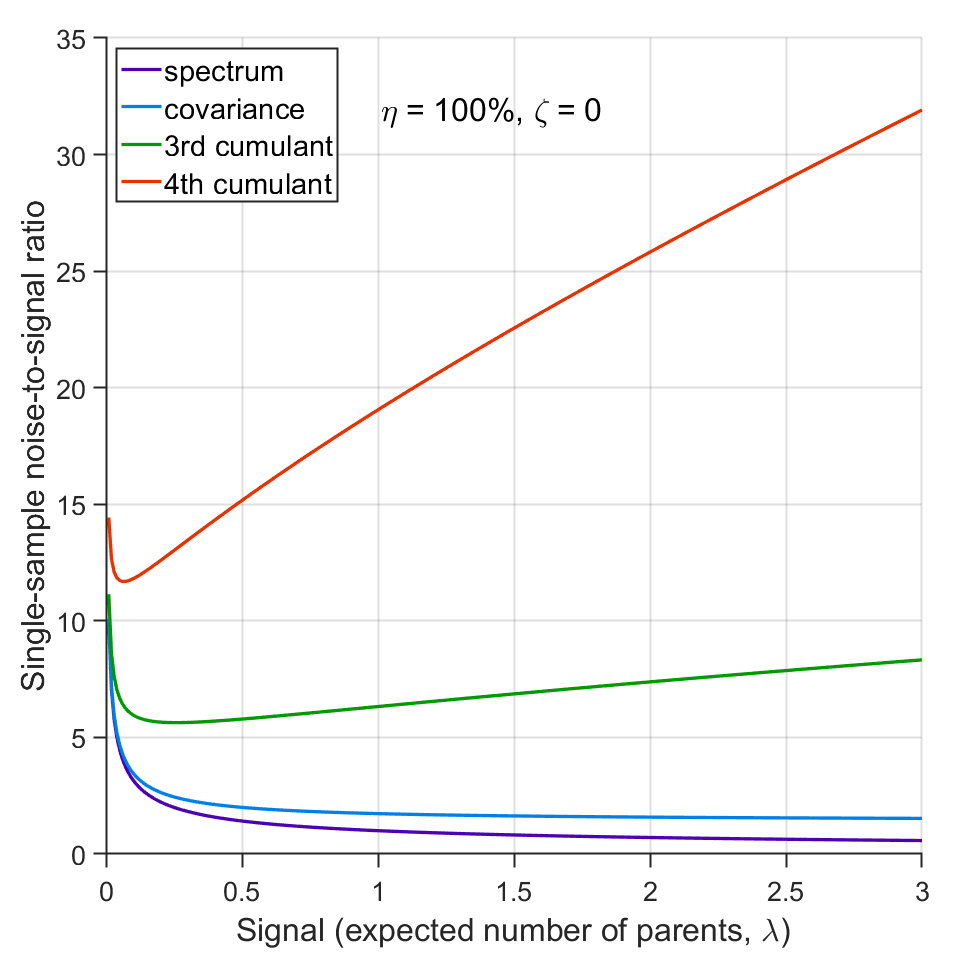}
  \caption{\label{fig_noiseLambda} \textbf{Cumulant noise as the function of the number of parent objects under the ideal conditions.} Unlike for the first and second cumulants, the noise for the higher cumulants is minimal at low counting rates.}
\end{figure}

It is instructive first to look at the results under the ideal conditions of full detection efficiency and no background. Substituting \(\eta = 100\%\) and \(\zeta = 0\) into the equations in Section \ref{summary} we find that there is no contribution from the uncorrelated parts and the expressions for the noise-to-signal ratio (N/S) calculated from Eq.~\ref{NS} are relatively simple functions of \(\lambda\).
These functions are plotted in Fig.~\ref{fig_noiseLambda} for \(N = 1\).

With increasing counting rate, i.e. increasing \(\lambda\), the noise of the first cumulant approaches zero as \(1/\sqrt{\lambda}\), which reflects the well-known fact that a high counting rate is always advantageous in collecting 1D spectra.

The N/S of the second cumulant (covariance) approaches a constant value of \(\sqrt{2}\) with an increasing counting rate. This tells us that for covariance mapping there is little advantage in increasing \(\lambda\) beyond 1 or 2, unless we want to accommodate weak and strong features on the same map. (In fact a high counting rate exacerbates map distortions due to fluctuations in experimental conditions, which induce common-mode fragment correlations \cite{frasinski2013dynamics, kornilov2013coulomb}.)

Cumulants higher than the second one have N/S increasing at high counting rates due to the higher powers of \(\lambda\) present in the expressions for \(\text{var}(\widehat{\varkappa_n})\). Therefore, for \(n \geq 3\) there is an optimal counting rate at low values of \(\lambda\) as the minima of the green and orange curves show in Fig.~\ref{fig_noiseLambda}.

\subsection{Reduced detection efficiency}

\begin{figure}[b]  
  \includegraphics[width=\linewidth]{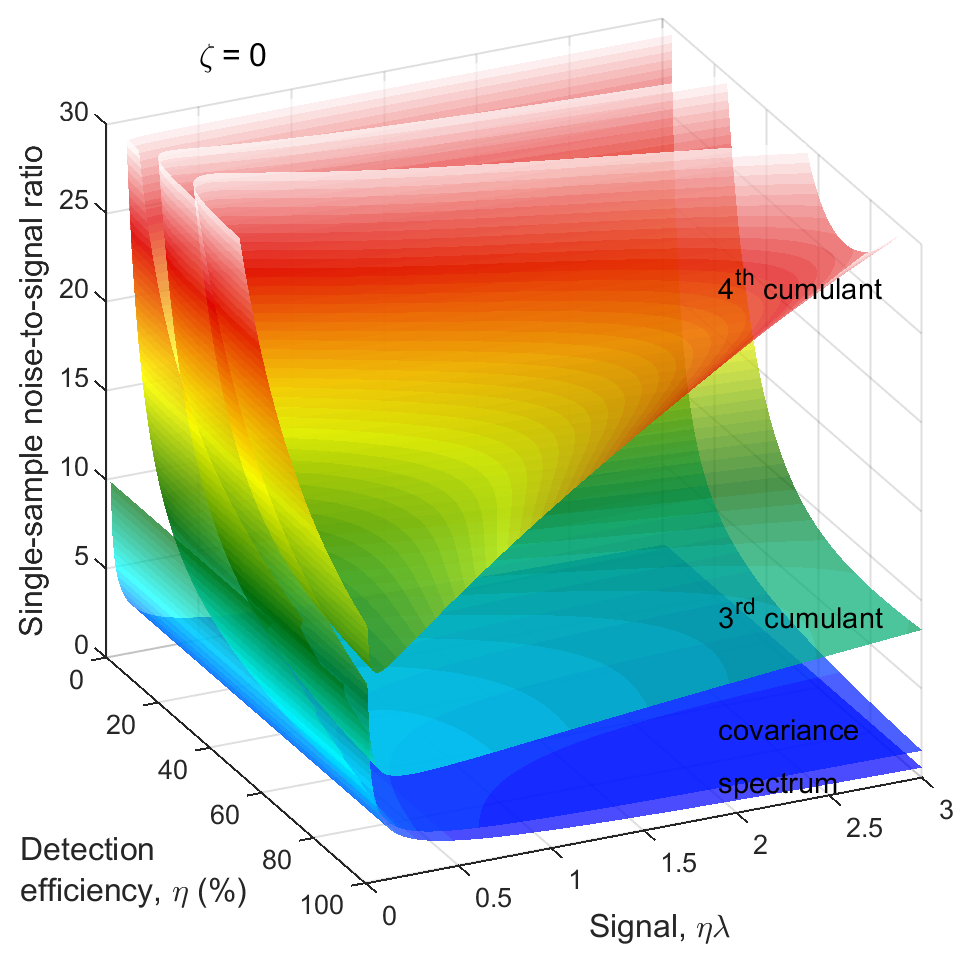}  
  \caption{\label{fig_noiseEta} \textbf{Cumulant noise as in Fig.~\ref{fig_noiseLambda} but resolved for \(\eta\).} When detection efficiency is reduced to about 50\%, there is only a modest increase in the noise.}
\end{figure}

To assess how a reduced detection efficiency affects the noise, we plot N/S as a function of \(\eta\) and \(\eta \lambda\), as shown in Fig.~\ref{fig_noiseEta}. The reason for choosing the latter argument rather than just \(\lambda\) is that \(\eta \lambda\) is the mean number of only the detected fragments, which is what is observed experimentally on 1D spectra. For simplicity, it is assumed that \(\eta\) is the same for every fragment.

As expected, the noise of the 2\textsuperscript{nd} and higher cumulants substantially increases when the detection efficiency is very low. However, when the detection efficiency is reduced only moderately, to around 50\%, the increase in the noise is also moderate, even for the 4\textsuperscript{th} cumulant. Since 50\% detection efficiency is within the reach of modern particle detectors, it makes cumulant mapping a feasible proposition.

\subsection{Background fragments}

\begin{figure}[t]  
  \includegraphics[width=\linewidth]{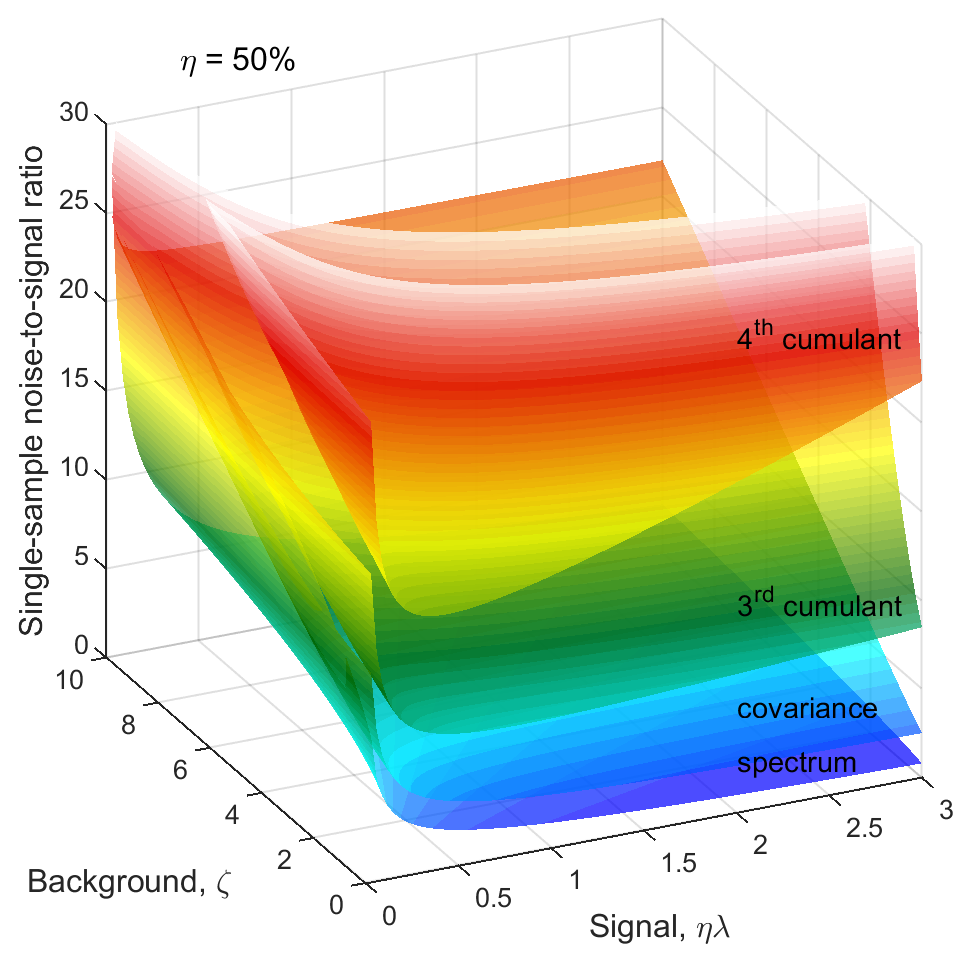}  
  \caption{\label{fig_noiseZeta} \textbf{Cumulant noise as in Fig.~\ref{fig_noiseLambda} but at a reduced \(\eta\) and resolved for \(\zeta\).} The noise increases with increasing background, especially for the higher cumulants and higher counting rates.}
\end{figure}

The next correction to the ideal conditions worth considering is a background of uncorrelated fragments (magenta circles in Fig.~\ref{fig_principle}). This is done using a realistic value of \(\eta = 50\%\) and plotting N/S as a function of \(\zeta\) and \(\eta \lambda\) in Fig.~\ref{fig_noiseZeta}. This choice of arguments means that they are proportional, respectively, to the relative background level and the height of a peak on a 1D spectrum, as shown in Fig.~\ref{fig_kappa1}.

For simplicity, we assume the same background level, \(\zeta\), for each kind of fragment, which makes the noise of higher cumulants to grow faster with increasing \(\zeta\) than the lower ones because of the higher powers of \(\zeta\) present in the expressions for \(\text{var}(\widehat{\varkappa_n})\). In some experiments this may be an over-pessimistic assumption because some of the \(Z, Y, X, \text{ or } W\) fragments may experience little or no background at all. The code given in Supplemental Material accepts \(\zeta\) and \(\eta\) tailored to each kind of fragment.

The optimum counting rate is broadly the same as for no background shown in Fig.~\ref{fig_noiseEta}. With an increasing background level, however, the optima for the higher cumulants shift to even lower counting rates.

\subsection{Number of samples needed}

When planning an experiment, the calculated noise is used to estimate the number of samples, \(N\), needed to suppress N/S to an acceptable level. We can use Eq.~\ref{NS} to calculate \(N\) for a fixed noise level, e.g. N/S = 0.1. Taking \(\eta = 50\%\), the result is shown in Fig.~\ref{fig_samplesZeta} on a logarithmic scale. The need to reduce the counting rate is clearly visible for the higher cumulants, especially at high background levels. The optimal \(\eta \lambda\) for the 3\textsuperscript{rd} and 4\textsuperscript{th} cumulants is below 0.05 at \(\zeta > 5\), which means that the observed fragments should be detected in less than 1 in 20 single-sample spectra. Such a low counting rate is comparable to the requirement of coincidence experiments. Unlike coincidences, however, cumulant mapping can accommodate higher counting rates if necessary, albeit at an increased noise.

\subsection{Practical implications}

\begin{figure}[t]  
  \includegraphics[width=\linewidth]{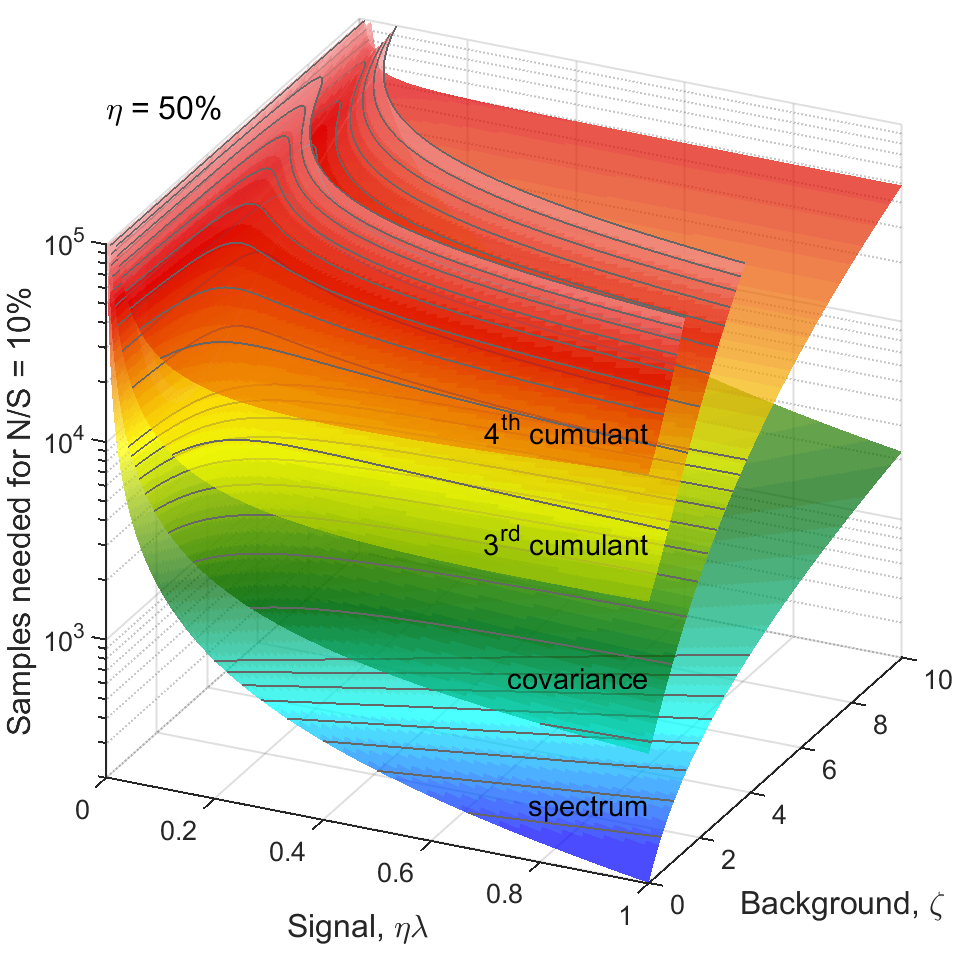}  
  \caption{\label{fig_samplesZeta} \textbf{Number of samples needed for a fixed noise-to-signal ratio.} In realistic experimental conditions about 10\textsuperscript{5} samples are needed to build a clear 4\textsuperscript{th} cumulant map.}
\end{figure}

Fig.~\ref{fig_samplesZeta} tells us that at least on the order of 10\textsuperscript{5} samples will be needed to obtain a good 4\textsuperscript{th} cumulant map. If we want to complete data collections in 15--20 minutes, then the sampling rate should be at least 100 Hz, and 1 kHz or more is desirable.

Such sampling rates are now routinely available from femtosecond lasers and becoming available from XFELs \cite{huang2021features}. For example, the LCLS-II XFEL will be operating at up to 1 MHz repetition rate, enabling researchers to probe over 10\textsuperscript{9} samples in a single experimental run. In principle, such a large number of samples makes it possible to build clear cumulant maps of even higher order than the 4\textsuperscript{th} one. In practice, however, the data acquisition speed is likely to be the limiting factor, since every single-sample spectrum needs to be recorded.

Cumulant mapping can significantly enhance the conventional mass spectrometry, whose main application is in identifying large biomolecules. The conventional approach is to obtain a high-quality mass spectrum of fragments and search for a match in a large database of molecular spectra. Therefore, the development of commercial spectrometers is driven towards the high mass resolution at the expense of the repetition rate, which is normally below 1 Hz.

Recently covariance mapping has been successfully applied to analyse mass spectra obtained from a commercial spectrometer \cite{driver2020two, driver2021two}. Rather than relying on the high mass resolution, the technique resolves the spectra in the second dimension and partially reconstructs the parent objects on a 2D map. Such a reconstruction allows some parent identifications that would be impossible using only a 1D spectrum of any quality. Clearly, this technique can utilise higher-cumulant mapping to perform a more complete parent reconstruction. Since many samples are needed to build cumulant maps, the spectrometer should operate at high repetition rates, for example, by employing the time-of-flight technique.

The volume of a multi-dimensional cumulant map can be very large and cumbersome to explore. In the simplest approach, cross-sections and projections of the map can be used to visualise the reconstructed molecules \cite{frasinski1991multiphoton}. If the locations of the reconstructed molecules are to be found, computational methods of artificial intelligence can be used to discover and identify them.

Cumulant mapping spectrometry can be combined with laser-induced fragmentation. On one hand, such combination allows us to elucidate the dynamics of molecular ionization and fragmentation \cite{allum2021multi}, on the other hand, it makes it possible to tune the fragmentation to specific bonds in large biomolecules \cite{ayers2022covariance}.

\section{Conclusions and outlook}

Cumulant mapping forms a firm theoretical basis for the concept of `multi-fold covariance'. The derived formulae enable the experimentalist to assess quantitatively the feasibility of studying multiple correlations in a fragmentation experiment. The key requirements are detection efficiency of around 50\%, and a sufficient number of samples, which in practice translates to sampling at high repetition rates.

Studies of molecular fragmentations induced by femtosecond or x-ray lasers are obvious areas for applying cumulant mapping. The technique can be used to extend the conventional mass spectrometry to multiple dimensions and substantially enhance its selectivity. Extension to particles other than electrons or ions should be straightforward. In particular, photons in a wide spectral range from near infrared to hard gamma rays are the promising candidates.

Since cumulant mapping relies on the Poisson distribution of the samples, in principle, it is applicable to any repetitive Poissonian process. For example, the neuronal spike trains closely follow Poisson point processes \cite{gardella2019modeling, campo2020inferring}. It could be speculated that cumulants represent the high-level brain functions emerging from correlations in low-level neuronal activity.

\bibsection  

\section*{Appendix: Variance calculations}

\appendix*

\renewcommand{\theequation}{\arabic{equation}}

Variance of \(\widehat{\varkappa_n}\) is needed to estimate the number of samples that have to be collected to measure the cumulant with given noise-to-signal ratio. Since only an estimate is needed, I will use approximations in deriving some of the formulae.

\subsection{Useful formulae}

The expected value of random variables, or functions of random variables, is distributive over summation:
\begin{equation}
  \langle A + B \rangle = \langle A \rangle + \langle B \rangle.
  \label{exp-sum}
\end{equation}
This is generally not true for multiplication, unless \(A\) and \(B\) are independent:
\begin{equation}
  \langle AB \rangle = \langle A \rangle \langle B \rangle
    + \text{cov}(A, B).
  \label{exp-prod}
\end{equation}

Their variances and the covariance can be expressed in terms of expected values:
\begin{equation}
  \text{var}(A)
  = \langle (A - \langle A \rangle)^2 \rangle
  = \langle A^2 \rangle - \langle A \rangle^2,
  \label{var-def}
\end{equation}
\begin{multline}
 \text{cov}(A, B) 
  = \langle (A - \langle A \rangle) \, (B - \langle B \rangle) \rangle \\
  = \langle AB \rangle - \langle A \rangle \langle B \rangle.
 \label{cov-def}
\end{multline}

An expected value of a sample average is equal to the expected value of the whole population:
\begin{equation}
  \big\langle \overline{A} \big\rangle = \langle A \rangle,
  \label{exp-ave}
\end{equation}
but the variance and covariance are scaled down by the number of samples \cite[Ref.][Eqs. 10.7, 10.22]{kendall94vol1}:
\begin{align}
  \text{var} \big(\overline{A} \big)
  &= \frac{1}{N} \big(\langle A^2 \rangle - \langle A \rangle^2 \big),
  \label{var-ave} \\
  \text{cov} \big(\overline{A}, \overline{B} \big)
  &= \frac{1}{N} \big(\langle AB \rangle
   - \langle A \rangle \langle B \rangle  \big).
  \label{cov-ave}
\end{align}

Since covariance is a linear function, the covariance of a random variable \(A\) with the sum of several other random variables \(X_1, X_2, X_3, ...\) is the sum of the covariances:
\begin{equation}
  \text{cov} \Big(A, \sum_i X_i \Big) = \sum_i \text{cov} (A, X_i).
  \label{cov-sum}
\end{equation}

Whereas variance is a nonlinear function, it can be expanded as follows:
\begin{multline}
  \text{var} \Big( \sum_i X_i \Big)
  = \text{cov} \Big( \sum_i X_i, \sum_j X_j \Big) \\
  = \sum_i \sum_j \text{cov}(X_i, X_j) \\
  = \sum_i \text{var}(X_i) + 2 \sum_i \sum_{j<i} \text{cov}(X_i, X_j).
  \label{var-sum}
\end{multline}

The variance and covariance of products of random functions can be calculated using the delta method. If \(F(\bm{X})\) and \(G(\bm{X})\) are functions of a random vector \(\bm{X} = [X_1, X_2, X_3, ...]\), then the Taylor expansion of \(F\) and \(G\) gives \cite[Ref.][Eqs. 10.12, 10.13]{kendall94vol1}
\begin{multline}
  \text{var}\big( F(\bm{X}) \big)
  \approx \sum_i 
    \left. \frac{\partial F}{\partial X_i} \right\rvert_{\bm{\xi}}^{\: 2}
    \text{var}(X_i) \\
  + \sum_i \sum_{j \neq i}
    \left. \frac{\partial F}{\partial X_i} \right\rvert_{\bm{\xi}}
    \left. \frac{\partial F}{\partial X_j} \right\rvert_{\bm{\xi}}
    \text{cov}(X_i, X_j),
  \label{var-delta}
\end{multline}
\begin{multline}
  \text{cov}\big( F(\bm{X}), G(\bm{X}) \big)
  \approx \sum_i
    \left. \frac{\partial F}{\partial X_i} \right\rvert_{\bm{\xi}}
    \left. \frac{\partial G}{\partial X_i} \right\rvert_{\bm{\xi}}
    \text{var}(X_i) \\
  + \sum_i \sum_{j \neq i}
    \left. \frac{\partial F}{\partial X_i} \right\rvert_{\bm{\xi}}
    \left. \frac{\partial G}{\partial X_j} \right\rvert_{\bm{\xi}}
    \text{cov}(X_i, X_j),
  \label{cov-delta}
\end{multline}
where the partial derivatives are evaluated at a particular value of \(\bm{X}\), which in this work is set to its expected value \(\bm{\xi} = \langle \bm{X} \rangle\).

To calculate the variance of the product of two random variables, I make the following choice for Eq.~\ref{var-delta}:
\begin{equation*}
  F = AB, \;\; \bm{X} = [A, B], \;\;
    \bm{\xi} = \big[ \langle A \rangle, \langle B \rangle \big],
\end{equation*}
which allows me to calculate the partial derivatives:
\begin{equation*}
  \left. \frac{\partial F}{\partial A} \right\rvert_{\bm{\xi}}
    = \langle B \rangle, \;\;
  \left. \frac{\partial F}{\partial B} \right\rvert_{\bm{\xi}}
    = \langle A \rangle.
\end{equation*}
Substituting these intermediate results into Eq.~\ref{var-delta} gives us the required formula:
\begin{multline}
  \text{var}(AB) \approx \langle A \rangle^2 \text{var}(B)
  + \langle B \rangle^2 \text{var}(A) \\
  + 2 \, \text{cov}(A, B),
  \label{var-prod}
\end{multline}
including the special case when \(B = A\):
\begin{equation}
  \text{var}(A^2) \approx 4 \langle A \rangle^2 \text{var}(A).
  \label{var-squa}
\end{equation}

In a similar way if I choose the following for Eq.~\ref{cov-delta}:
\begin{gather*}
  F = AB, \;\; G = CD, \;\; \bm{X} = [A, B, C, D], \\
  \bm{\xi} = \big[\langle A \rangle, \langle B \rangle,
    \langle C \rangle, \langle D \rangle \big].
\end{gather*}
and calculate the other two non-zero partial derivatives
\begin{equation*}
  \left. \frac{\partial G}{\partial C} \right\rvert_{\bm{\xi}}
    = \langle D \rangle, \;\;
  \left. \frac{\partial G}{\partial D} \right\rvert_{\bm{\xi}}
    = \langle C \rangle,
\end{equation*}
I obtain the formula for the covariance of products:
\begin{multline}
  \text{cov}(AB, CD) \approx
    \langle A \rangle \langle C \rangle \text{cov}(B, D)
  + \langle A \rangle \langle D \rangle \text{cov}(B, C) \\
  + \langle B \rangle \langle C \rangle \text{cov}(A, D)
  + \langle B \rangle \langle D \rangle \text{cov}(A, C).
  \label{cov-prod}
\end{multline}

When estimating experimental cumulants we use sample averages, \(\overline{\bm{X}}\), to center random variables. In theoretical estimations the expected values, \(\langle \bm{X} \rangle\), can be used instead, which significantly simplifies the calculations. These two versions of the cumulant estimators are asymptotically equal when the number of samples tends to infinity:
\begin{equation}
  \widehat{\varkappa_n}(\bm{X} - \overline{\bm{X}})
  \simeq \widehat{\varkappa_n}(\bm{X} - \langle \bm{X} \rangle).
  \label{estim-approx}
\end{equation}
This approximation makes no difference when calculating the expected cumulant values, or the variance of the first and the second cumulants. For the third and higher cumulants the experimental variance is a little lower than the theoretical one. This is because the sample average contributes additional negative terms in powers of 1/\textit{N} expansion to the variance \cite[Ref.][Example 10.3]{kendall94vol1}. By using the right hand side of Eq.~\ref{estim-approx}, I effectively estimate deviations from the true, unknown value of the cumulant, rather than the noise on the map.

Another approximation I use, also affecting only the variance of the third and higher cumulants, is the neglect of the residual correlations that are discussed following Eq.~\ref{vanish}.

\subsection{1D spectrum}

The calculation of the variance quoted in Section \ref{summ1D} is straightforward:
\begin{multline*}
  \text{var}\big( \widehat{\varkappa_1} \big)
  \overset{\text{\ref{kappa-1-estim}}}{=}
    \text{var} \big( \overline{Z} \big)
  \overset{\text{\ref{var-def},\ref{var-ave}}}{=}
    \frac{1}{N} \big\langle (Z - \langle Z \rangle)^2  \big\rangle \\
  \overset{\text{\ref{z0-y0},\ref{z0-y0-x0}}}{=}
    \frac{1}{N} \big\langle (s + z)^2 \big\rangle
  \overset{\text{\ref{vanish}}}{=}
     \frac{1}{N} \big( \langle s^2 \rangle + \langle z^2 \rangle \big),
\end{multline*}
where the numbers above equality signs refer to the equations used. The last equality relies on the independence of \(s\) and \(z\), which is always possible to satisfy having only one variable \(Z\).

\subsection{2D covariance map}

To start variance calculations we write the estimator of the second cumulant in terms of mean-centered variables:
\begin{multline*}
  \widehat{\varkappa_2}
  \overset{\text{\ref{estim-approx}}}{\simeq}
    \overline{(Z - \langle Z \rangle)(Y - \langle Y \rangle)}
  \overset{\ref{z0-y0},\ref{z0-y0-x0}}{=}
    \overline{(s + z)(s + y)} \\
  = \overline{s^2} + \overline{s(z + y)} + \overline{zy},
\end{multline*}
and expand the variance of the sum:
\begin{multline*}
  \text{var}\big( \widehat{\varkappa_2} \big)
  \overset{\text{\ref{var-sum}}}{=}
    \text{var}\big( \overline{s^2} \big)
  + \text{var}\big( \overline{s(z + y)} \big)
  + \text{var}\big( \overline{zy} \big) \\
  + \text{covariance terms}.
\end{multline*}

We calculate the variance terms one by one:
\begin{equation*}
  \text{var}\big( \overline{s^2} \big)
  \overset{\text{\ref{var-ave}}}{=}
    \frac{1}{N} \big( \langle s^4 \rangle - \langle s^2 \rangle^2 \big),
\end{equation*}
\begin{multline*}
  \text{var}\big( \overline{s(z + y)} \big)
  \overset{\text{\ref{var-ave}}}{=}
    \frac{1}{N} \big( \langle s^2(z + y)^2 \rangle
  - \langle s(z + y) \rangle^2 \big) \\
  \overset{\text{\ref{exp-sum},\ref{exp-prod}}}{=}
    \frac{1}{N} \Big( \langle s^2 \rangle
    \big(\langle z^2 \rangle + \langle y^2 \rangle
  + 2 \langle z \rangle \langle y\rangle \big) \\
  - \big(\langle s \rangle (\langle z \rangle
    + \langle y \rangle) \big)^2 \Big) \\
  \overset{\text{\ref{vanish}}}{=}
  \frac{1}{N} \Big( \langle s^2 \rangle
    \big(\langle z^2 \rangle + \langle y^2 \rangle \big) \Big),
\end{multline*}
\begin{equation*}
  \text{var}\big( \overline{zy} \big)
  \overset{\text{\ref{var-ave}}}{=}
    \frac{1}{N} \big( \langle z^2 y^2 \rangle - \langle zy \rangle^2 \big)
  \overset{\text{\ref{exp-prod},\ref{vanish}}}{=}
    \frac{1}{N} \langle z^2 \rangle \langle y^2 \rangle.
\end{equation*}

Note a useful simplification of Eq.~\ref{vanish}: a term vanishes if it has a factor linear in the expected value of a mean-centered variable. For this reason all the covariance terms are zero, for example
\begin{multline*}
  \text{cov} \big(\overline{s^2}, \overline{s(z + y)} \big)
  \overset{\text{\ref{cov-ave}}}{=}
    \frac{1}{N} \big(\langle s^2 s(z + y) \rangle
  - \langle s^2 \rangle \langle s(z + y) \rangle  \big) \\
  \overset{\text{\ref{exp-sum},\ref{exp-prod}}}{=}
   \frac{1}{N} \Big( \langle s^3 \rangle
    \big(\langle z \rangle + \langle y \rangle \big)
  - \langle s^2 \rangle \langle s \rangle
    \big(\langle z \rangle + \langle y \rangle \big) \Big)
  \overset{\text{\ref{vanish}}}{=} 0.
\end{multline*}

Gathering all the variance terms we obtain the formula for \(\text{var}\big( \widehat{\varkappa_2} \big)\) given in Section \ref{summ2D}.

\subsection{3D cumulant map}

The calculation of the third cumulant estimator is similar to the second one:
\begin{multline*}
  \widehat{\varkappa_3}
  \overset{\text{\ref{estim-approx}}}{\simeq}
    \overline{(Z - \langle Z \rangle)(Y - \langle Y \rangle)
      (X - \langle X \rangle)} \\
  \overset{\ref{z0-y0},\ref{z0-y0-x0}}{=}
    \overline{(s + z)(s + y)(s + x)} \\
  = \overline{s^3} + \overline{s^2 (z+y+x)} + \overline{s(zy+zx+yx)}
    + \overline{zyx}.
\end{multline*}

Expanding the variance of the sum we find that the covariance terms vanish again and we need to calculate only variances:
\begin{equation*}
  \text{var}\big( \overline{s^3} \big)
  \overset{\text{\ref{var-ave}}}{=}
    \frac{1}{N} \big( \langle s^6 \rangle - \langle s^3 \rangle^2 \big),
\end{equation*}
\begin{multline*}
  \text{var}\big( \overline{s^2 (z + y + x)} \big) \\
  \overset{\text{\ref{var-ave}}}{=}
    \frac{1}{N} \big( \langle s^4(z + y + x)^2 \rangle
  - \langle s(z + y + x) \rangle^2 \big) \\
  \overset{\text{\ref{exp-prod}}}{=}
    \frac{1}{N} \Big( \langle s^4 \rangle
    \Big\langle \Big( \sum^3 z \Big)^2 \Big\rangle
  - \langle s \rangle^2
    \Big\langle \sum^3 z \Big\rangle^2 \Big) \\
  \overset{\text{\ref{exp-sum},\ref{exp-prod},\ref{vanish}}}{=}
    \frac{1}{N} \langle s^4 \rangle \sum^3 \langle z^2 \rangle,
\end{multline*}
\begin{multline*}
  \text{var}\big( \overline{s(zy + zx + yx)} \big) \\
  \overset{\text{\ref{var-ave}}}{=}
    \frac{1}{N} \big( \langle s^2 (zy + zx + yx)^2 \rangle
  - \langle s(zy + zx + yx) \rangle^2 \big) \\
  \overset{\text{\ref{exp-prod}}}{=}
    \frac{1}{N} \Big( \langle s^2 \rangle
    \Big\langle \Big( \sum^3 zy \Big)^2 \Big\rangle
  - \langle s \rangle^2
    \Big\langle \sum^3 zy \Big\rangle^2 \Big) \\
  \overset{\text{\ref{exp-sum},\ref{exp-prod},\ref{vanish}}}{=}
    \frac{1}{N} \langle s^2 \rangle
    \sum^3 \langle z^2 \rangle \langle y^2 \rangle,
\end{multline*}
\begin{multline*}
  \text{var}\big( \overline{zyx} \big)
  \overset{\text{\ref{var-ave}}}{=}
    \frac{1}{N} \big( \langle z^2 y^2 x^2 \rangle
    - \langle zyx \rangle^2 \big) \\
  \overset{\text{\ref{exp-prod},\ref{vanish}}}{=}
    \frac{1}{N}
    \langle z^2 \rangle \langle y^2 \rangle \langle x^2 \rangle.
\end{multline*}

Gathering all the variance terms we obtain the formula for \(\text{var}\big( \widehat{\varkappa_3} \big)\) given in Section \ref{summ3D}.

\subsection{4D cumulant map}

The calculation for the fourth cumulant is more laborious because not only the formula for the estimator is longer but also some of the covariance terms in the variance expansion do not vanish. While using more advanced tools, such as \textit{k}-statistics \cite{kendall94vol1}, may shorten the calculations, I shall continue with the same method as before: 
\begin{align*}
  \widehat{\varkappa_4}
  &\overset{\text{\ref{estim-approx}}}{\simeq}
     \overline{(Z - \langle Z \rangle)(Y - \langle Y \rangle)
      (X - \langle X \rangle)(W - \langle W \rangle)} \\
  &- \sum^3 \widehat{\varkappa_2}(Z, Y) \: \widehat{\varkappa_2}(X, W) \\
  &\overset{\ref{z0-y0},\ref{z0-y0-x0}}{=}
     \overline{(s + z)(s + y)(s + x)(s + w)} \\
  &- \sum^3 \Big(\overline{s^2} + \overline{s(z+y)} + \overline{zy} \Big)
           \Big(\overline{s^2} + \overline{s(x+w)} + \overline{xw} \Big) \\
  &= \overline{s^4} + \sum^4 \overline{s^3 z} + \sum^6 \overline{s^2 zy}
   + \sum^4 \overline{s zyx} + \overline{zyxw} \\
  &- 3 \, \overline{s^2}^{\,2}
   - 3 \, \overline{s^2} \sum^4 \overline{sz}
   - \overline{s^2} \sum^6 \overline{zy} \\
  &- 2 \sum^6 \overline{sz} \; \overline{sy}
   - \sum^{12} \overline{sz} \; \overline{yx}
   - \sum^3 \overline{zy} \; \overline{xw}.
\end{align*}

To calculate the variance of this estimator we apply Eq.~\ref{var-sum} and evaluate the variance of the first two terms:
\begin{equation*}
  \text{var}\big( \overline{s^4} \big)
  \overset{\text{\ref{var-ave}}}{=}
    \frac{1}{N} \big( \langle s^8 \rangle - \langle s^4 \rangle^2 \big),
\end{equation*}
\begin{multline*}
  \text{var}\Big( \sum^4 \overline{s^3 z} \Big)
  \overset{\ref{var-sum},\ref{vanish}}{=}
    \sum^4 \text{var}\big( \overline{s^3 z} \big) + 0 \\
  \overset{\text{\ref{var-ave},\ref{vanish}}}{=}
    \frac{1}{N} \Big( \sum^4 \langle s^6 z^2 \rangle - 0^2 \Big)
  \overset{\text{\ref{exp-prod}}}{=}
    \frac{1}{N} \langle s^6 \rangle \sum^4 \langle z^2 \rangle,
\end{multline*}
and continue applying Eqs. \ref{var-sum}, \ref{var-ave}, \ref{exp-prod}, and \ref{vanish} to the next three terms:
\begin{equation*}
  \text{var}\Big( \sum^6 \overline{s^2 zy} \Big)
  = \frac{1}{N} \langle s^4 \rangle \sum^6
    \langle z^2 \rangle \langle y^2 \rangle,
\end{equation*}
\begin{equation*}
  \text{var}\Big( \sum^4 \overline{szyx} \Big)
  = \frac{1}{N} \langle s^2 \rangle \sum^4
    \langle z^2 \rangle \langle y^2 \rangle \langle x^2 \rangle,
\end{equation*}
\begin{equation*}
  \text{var}\big( \overline{zyxw} \big)
  = \frac{1}{N} \langle z^2 \rangle \langle y^2 \rangle
    \langle x^2 \rangle \langle w^2 \rangle.
\end{equation*}

To calculate the variance of the next three terms we can use Eqs. \ref{var-prod} and \ref{var-squa} for the variance of a product of two variables:
\begin{multline*}
  \text{var}\Big( {-3} \, \overline{s^2}^{\,2} \Big)
  = 9 \, \text{var}\Big( \overline{s^2} \; \overline{s^2} \Big)
  \overset{\ref{var-squa}}{\approx}
    36 \,\big\langle \overline{s^2} \big\rangle^2
    \,\text{var}\big( \overline{s^2} \big) \\
  \overset{\ref{exp-ave},\ref{var-ave}}{=}
    \frac{36}{N} \langle s^2 \rangle^2
    \big( \langle s^4 \rangle - \langle s^2 \rangle^2 \big),
\end{multline*}
\begin{multline*}
  \text{var}\Big( {-3} \, \overline{s^2} \; \sum^4 \overline{sz} \Big)
  \overset{\ref{var-prod},\ref{vanish}}{\approx}
    9 \,\big\langle \overline{s^2} \big\rangle^2
    \,\text{var}\Big( \sum^4 \overline{sz} \Big) + 0 + 2 \times 0 \\
  \overset{\ref{exp-ave},\ref{var-sum},\ref{vanish}}{=}
    9 \,\langle s^2 \rangle^2
    \,\Big( \sum^4 \text{var} \big( \overline{sz} \big) + 0 \Big) \\
  \overset{\ref{var-ave},\ref{vanish}}{=}
    \frac{9}{N} \langle s^2 \rangle^2
    \sum^4 \big( \langle s^2 z^2 \rangle - 0^2 \big) \\
  \overset{\ref{exp-prod}}{=}
    \frac{9}{N} \langle s^2 \rangle^3
    \sum^4 \langle z^2 \rangle,
\end{multline*}
and in a very similar way we find
\begin{equation*}
  \text{var}\Big( {-\overline{s^2}} \, \sum^6 \overline{zy} \Big)
  \approx \frac{1}{N} \langle s^2 \rangle^2
    \sum^6 \langle z^2 \rangle \langle y^2 \rangle.
\end{equation*}

When we proceed to calculate the variances of the last line of \(\widehat{\varkappa_4}\), we notice that all terms in Eq.~\ref{var-prod} vanish. Therefore all remaining variance terms are zero.

To calculate the covariance terms of \(\widehat{\varkappa_4}\), we refer to Eq.~\ref{cov-ave} and note that the covariance vanishes unless both \(A\) and \(B\) have the same subset of the \(z,y,x,w\) variables. Therefore, the only non-zero covariance terms are:
\begin{multline*}
  2 \,\text{cov}\Big( \overline{s^4}, {-3} \, \overline{s^2}^{\,2} \Big)
  = {-6} \, \text{cov}
    \big( \overline{s^4}, \overline{s^2} \; \overline{s^2} \big) \\
  \overset{\ref{cov-prod},\ref{vanish}}{\approx}
    {-6} \times 2 \,\big\langle \overline{s^2} \big\rangle^2
    \,\text{cov}\big( \overline{s^4}, \overline{s^2} \big) + 0 \\
  \overset{\ref{exp-ave},\ref{cov-ave}}{=}
    {-\frac{12}{N}} \langle s^2 \rangle
    \Big( \langle s^6 \rangle
    - \langle s^4 \rangle \langle s^2 \rangle \Big),
\end{multline*}
\begin{multline*}
  2 \,\text{cov}\Big( \sum^4 \overline{s^3 z},
    {-3} \, \overline{s^2} \sum^4 \overline{sz} \Big) \\
  \overset{\ref{cov-sum},\ref{vanish}}{=}
    {-6} \sum^4 \text{cov}
    \big( \overline{s^3 z}, \overline{s^2} \; \overline{sz} \big) + 0\\
  \overset{\ref{cov-prod},\ref{vanish}}{\approx}
    {-6} \sum^4 \Big( \big\langle \overline{s^2} \big\rangle
    \,\text{cov}\big( \overline{s^3 z}, \overline{s z} \big) + 0 \Big) \\
  \overset{\ref{exp-ave},\ref{cov-ave},\ref{vanish}}{=}
    {-\frac{6}{N}} \langle s^4 \rangle \langle s^2 \rangle
    \sum^4 \langle z^2 \rangle,
\end{multline*}
and in a very similar way
\begin{multline*}
  2 \,\text{cov}\Big( \sum^6 \overline{s^2 zy},
    {-\overline{s^2}} \sum^6 \overline{zy} \Big) \\
  \approx {-\frac{2}{N}} \langle s^2 \rangle^2
    \sum^6 \langle z^2 \rangle \langle y^2 \rangle.
\end{multline*}

Gathering all the variance and covariance terms gives us the formula for \(\text{var}\big( \widehat{\varkappa_4} \big)\) given in Section \ref{summ4D}.


\bibliography{Cumulant} 

\end{document}